\documentclass[nonacm=true]{acmart}

\long\def\COMMENT#1\ENDCOMMENT{\message{(Commented text...)}\par}

\restylefloat{table}
\usepackage{graphicx}
\usepackage{tabularx}

\AtBeginDocument{%
  \providecommand\BibTeX{{%
    \normalfont B\kern-0.5em{\scshape i\kern-0.25em b}\kern-0.8em\TeX}}}

\acmConference[ICAIF'20]{ACM AI in Finance}{October 15--16, 2020}{New York City, NY}
\acmBooktitle{}
\acmPrice{15.00}
\acmISBN{978-1-4503-XXXX-X/18/06}

\begin{document}

\title[Sentiment Anal. \& Volatility]{
A Sentiment Analysis Approach to the Prediction of Market Volatility}

\author{Justina Deveikyte}
\affiliation{%
  \institution{Dept. of Computer Science \& Information Systems\\
  Birkbeck, University of London}
  \city{London WC1E 7HX, UK}
  }
\email{jdevei01@dcs.bbk.ac.uk}

\author{Helyette Geman}
\orcid{}
\affiliation{%
  \institution{Dept. of Economics, Mathematics \& Statistics\\
  Birkbeck, University of London}
  \city{London WC1E 7HX, UK}
  }
\email{hgeman@hotmail.com}

\author{Carlo Piccari}
\orcid{}
\affiliation{%
  \institution{Dept. of Economics, Mathematics \& Statistics\\
  Birkbeck, University of London}
  \city{London WC1E 7HX, UK}
  }
\email{cpicca01@mail.bbk.ac.uk}

\author{Alessandro Provetti}
\orcid{0000-0001-9542-4110}
\affiliation{%
  \institution{Dept. of Computer Science \& Information Systems\\
  Birkbeck, University of London}
  \city{London WC1E 7HX, UK}
  }
\email{ale@dcs.bbk.ac.uk}

\renewcommand{\shortauthors}{Deveikyte et al.}

\begin{abstract}
Prediction and quantification of future volatility and returns play an important role in financial modelling, both in portfolio optimization and risk management. 
Natural language processing today allows to process news and social media comments to detect signals of investors' confidence.
We have explored the relationship between sentiment extracted from financial news and tweets and FTSE100 movements. 
We investigated the strength of the correlation between sentiment measures on a given day and market volatility and returns observed the next day. 
The findings suggest that there is evidence of correlation between sentiment and stock market movements: the sentiment captured from news headlines could be used as a signal to predict market returns; the same does not apply for volatility. 
Also, in a surprising finding, for the sentiment found in Twitter comments we obtained a correlation coefficient of -0.7, and p-value below 0.05, which indicates a strong negative correlation between positive sentiment captured from the tweets on a given day and the volatility observed the next day.
We developed an accurate classifier for the prediction of market volatility in response to the arrival of new information by deploying topic modelling, based on Latent Dirichlet Allocation, to extract feature vectors from a collection of tweets and financial news. 
The obtained features were used as additional input to the classifier. 
Thanks to the combination of sentiment and topic modelling our classifier achieved a directional prediction accuracy for volatility of 63\%.
\end{abstract}


\keywords{Sentiment Analysis, Topic Modelling, Stock Market Volatility, Correlation Analysis.
Machine Learning for pricing, trading, and portfolio management.
Models of financial behavior.}

\maketitle

\COMMENT
\section*{Motivation (not for inclusion)}
\emph{
Authors will also be asked to provide a paragraph, separate from the abstract, explaining “why this paper should be published at ICAIF.”} 

\textbf{
Our work presents a new software pipeline that combines recently-available tools from Natural Language Processing to create a robust predictor for Financial analysis.
Therefore, we believe that the ICAIF is the right venue where these results can be presented and discussed. 
Our work is in the direction pointed to by this new editorial project and we look forward to participate to the event.
}
\ENDCOMMENT
\section{Introduction}
Stock market returns and volatility prediction have attracted much attention from academia as well as the financial industry. 
But can stock market prices and volatility be predicted? 
Or at least, can they be predicted at some specific time? 
Measuring sentiment captured from online sources such as Twitter or financial news articles can be valuable in the development of trading strategies. 
In addition, sentiment captured from financial news can have some predictive power that can be harnessed by portfolio and risk managers. 

The results and conclusions from our analysis can be classified into three parts. 
First, correlations between sentiment scores and stock market returns were statistically significant for our headline dataset only. 
The results indicate a statistically significant negative correlation between negative news and the closing price of FTSE100 index (returns). 
The strongest correlation between sentiment and volatility measures was detected in our tweets dataset, while no correlation or weak correlation was found in headlines and news stories dataset. 
This can be explained by the fact that tweets can be timelier and more reactive to various events, whereas it takes much more time to publish articles, and the market functions according to the principle ``Buy on rumors,
sell on news.''

This paper is structured as follows. 
In Section \ref{sec:related} we review related research. 
In Section \ref{sec:data}, we describe data sources which have been used to calculate sentiment scores: Thomson Reuters, RavenPack and Twitter. 
In Section \ref{sec:correlation} we conduct a correlation analysis and Granger's causality test. 
In Section \ref{sec:lda}, we carry out additional experiments to access if topic modelling (or Latent Dirichlet Allocation) can be used to enhance the prediction accuracy of next days stock market directional volatility. 
In Section \ref{sec:results} and \ref{sec:conclusion} we present the results of the analysis and discuss future work.

\section{Related Works}\label{sec:related}
A growing number of research papers use NLP methods to access how sentiment of firm-specific news, financial reports, or social media impact stock market returns. 
An important early work (2007) by Tetlock \cite{Tet07} explores possible correlations between the media and the stock market using information from the Wall Street Journal and finds that high pessimism causes downward pressure on market prices. 
Afterwards, Tetlock et al. \cite{TSM08} uses a bag-of-words model to assess whether company financial news can predict a company’s accounting earnings and stock returns. 
The results indicate that negative words in company-specific news predict low firm earnings, although market prices tend under-react to the information entrenched in negative words. 

Bollen et al. \cite{Bol11} examined whether sentiment captured from Twitter feeds is correlated to the value of the Dow Jones Industrial Average Index (DJIA). 
They deployed OpinionFinder and Google-Profile of Mode States (GPOMS), opinion-tracking tools that measure mood in six dimensions (Calm, Alert, Sure, Vital, Kind, and Happy). 
The results let them to conclude that

\begin{quote}
    the accuracy of DJIA predictions can be significantly improved by the inclusion of specific public mood dimensions but not others.
\end{quote}

Loughran et al. \cite{LouMcD11} apply sentiment analysis to 10-K filings. Authors find that almost three-quarters of negative word counts in 10-K filings based on the Harvard dictionary are typically not negative in a financial context. 
To do so, they developed an alternative dictionary that better reflects sentiment in financial text.

A majority of the work in sentiment analysis seem to focus on predicting market prices or directional change. 
There are many examples of applying text mining to news data relating to the stock market with a particular emphasis on the prediction of market prices. 
However, only a limited number of research papers look into how financial news impacts stock market volatility. 

Kogan et al. \cite{KLRSS09} use Support Vector Machine (SVM) to predict the volatility of stock market returns. The results indicate that “text regression model predictions to correlate with true volatility nearly as well as historical volatility, and a combined model to perform even better”.

Mao et al. \cite{MCB11} use a wide range of news data and sentiment tracking measures to predict financial market values. The authors find that Twitter sentiment is a significant predictor of daily market returns, but after controlling for all other mood indicators including VIX, sentiment indicators are no longer statistically insignificant.

Similarly, Groß-Klußmann et al. \cite{HG11} find that the release of highly relevant news induces an increase in return volatility, with negative news having a greater impact than positive news.

Glasserman et al. \cite{GlaMam19} use an n-gram model to develop a methodology showing that unusual negative and positive news forecasts volatility at both the company-specific and aggregate levels. 
The authors find that an increase in the “unusualness" of news with negative sentiment predicts an increase in stock market volatility. Similarly, unusual positive news forecasts lower volatility. According to research findings, news is reflected in volatility more slowly at the aggregate than at the company-specific levell, in agreement with the effect of diversification. 

Calomiris et al. \cite{CM18} use news articles to develop a methodology to predict risk and return in stock markets in developed and emerging countries. 
Their results indicate that the topic-specific sentiment, frequency and unusualness of news text can predict future returns, volatility, and drawdowns. 

Similarly, Caporin et al. \cite{CP17} find that news-related variables can improve volatility prediction. 
Certain news topics such earning announcements and upgrades/downgrades are more relevant than other news variables in predicting market volatility. 

In a more recent study, Atkins et al. \cite{ANG18} use LDA and a simple Naive Bayes classifier to predict stock market volatility movements. 
The authors find that the information captured from news articles can predict market volatility more accurately than the direction the price movements. 
They obtained a 56\% accuracy in predicting directional stock market volatility on the arrival of new information.

Also Mahajan et al. \cite{MDH08} used LDA to identify topics of financial news and then to predict a rise or fall in the stock markets based on topics extracted from financial news. 
Their developed classifier achieved 60\% accuracy in predicting market direction.

Jiao et al. \cite{JW16} show that a high social media activity around a specific company predicts a significant increase in return volatility whereas attention from influential press outlets, e.g. the Wall Street Journal in fact is a  predictor of the opposite: a decrease in return volatility.

\section{Data}\label{sec:data}
For our research we decided to use three different data sets (tweets, news headlines, and full news stories) to analyse sentiment and compare the results. 
News headlines about FTSE100 companies were obtained from RavenPack. The dataset includes headlines as well as other metadata collected from 1 January 2019 to August 2019. 
News arrival is recorded with GMT time stamps up to a millisecond precision. 
In total we have 969,753 headlines for our analysis. 
The number of headlines during the weekends ranged from around 700 to 1,300 daily, while during normal working days the number of headlines often exceeded 5,000 per day.
We used the Eikon API%
\footnote{Please see \url{https://developers.refinitiv.com/eikon-apis/eikon-data-api}} 
to gather news stories about FTSE100 companies starting from April 2019 to the end of August 2019. 
Around 12,000 articles have been collected between April and August 2019. 

By using Twitter Streaming API, in total we collected 545,979 tweets during July–August 2019. 
For the purpose of this study and in order to avoid too generic tweets, we retained and mined only the so-called \texttt{``\$}cashtags'' that mentioned companies included in the FTSE100 index. 
The rationale for selecting certain hashtags relates back to the original aim of measuring sentiment of news related to FTSE100 companies rather than overall financial industry. 

For this project we decided to use FTSE100 index data. The FTSE100 index represents the performance of the largest 100 companies listed on the London Stock Exchange (LSE) with the highest market capitalization and is considered the best indicator of the health of the UK  stock market. 
The daily closing prices of FTSE100 index were obtained from the Reuters Eikon Platform, using their API. 

\emph{In addition, to assess the relationship between stock market movements and sentiment we computed daily market returns and  defined the return on day $t$ as the change in log Close from day $t-1$ expressed as}

\begin{equation}
  r_t=\log\frac{CLOSE_t}{CLOSE_{t-1}}
\end{equation}

The volatility of FTSE 100 is classically defined as:

\begin{equation}
  Vol = \sqrt{{\frac{1}{N}}\sum_{t+1}^N(r_t-\bar{r})^2} \cdot \sqrt{252}
\end{equation}
\subsection{The VADER Sentiment Intensity Analyzer}

According to VADER documentation, “VADER is a lexicon and rule-based sentiment analysis tool that is specifically attuned to sentiments expressed in social media”. 
This tool was developed by Hutto, C.J. and Gilbert, E.E. in 2014, but since than it underwent several improvements and updates.
The VADER sentiment analyser is extremely accurate when it comes to social media text because it provides not only positive and negative scores but also measures the intensity of the sentiment. 
Another advantage of using VADER is that it does not need training data as it uses human labelled sentiment lexicon and it works comparable fast.

The VADER lexicon was created manually, according to the documentation file which can be found on its GitHub site%
\footnote{Please see \url{https://github.com/cjhutto/vaderSentiment}}. 
Ten independent human raters annotated more than 9,000 token features on the following scale from -4 to +4:

\begin{itemize}
\item Extremely negative -4
\item Neutral score 0
\item Extremely positive +4
\end{itemize}

The positive, negative, and neutral scores are ratios for the proportions of text that fall in each category and should sum to 1.
The compound score is derived by summing the sentiment scores of each word in the lexicon, adjusted according to the rules, and then normalized to be between -1 (most extreme negative) and +1 (most extreme positive). 
This is the most useful metric if we want a single uni-dimensional measure of sentiment for a given sentence. 

The VADER sentiment analyser can handle negations, UTF-8 encoded emojis, as well as acronyms, slang and punctuation. 
Also, it takes punctuation into account by amplifying the sentiment score of the sentence proportional to the number of exclamation points and question marks ending the sentence. VADER first computes the sentiment score of the sentence. 
If the score is positive then VADER adds a certain empirically obtained quantity for every exclamation point (0.292) and question mark (0.18). 
Conversely, negative scores are subtracted.

\subsection{Aggregating News and Sentiment Scores}
In contrast to ﬁnancial stock data, news and tweets were available for each day, although the number of tweets and news was significantly lower during weekends and bank holidays. 
Not to lose that information, we decided to transfer the sentiment scores accumulated for non-trading days to the next nearest trading day. 
That is, the average news sentiment prevailing over weekend will be applied to the following Monday. 
The same logic holds for holidays. 

For our daily analysis, we aggregate sentiment scores captured from all tweets on day t to access its impact on the stock market performance in the coming ``$t+1$'' day. 
For instance, we aggregate sentiment captured from tweets on 10th July to analyse the correlation between the sentiment of that day and the coming day’s (11th July) market volatility and returns. 

We have adopted Gabrovšek et al. \cite{GAMG16} definition of the sentiment score \emph{Sentd:}

\begin{equation}
  Sent_d = \frac{{N_d(pos)}-{N_d(neg)}}{{N_d(pos)}+{N_d(neut)}+{N_d(neg)}+3}
  \label{eq:sentd}
\end{equation}

\medskip\noindent
where Nd(neg), Nd(neut), and Nd(pos) denote the daily volume of negative, neutral, and positive tweets and 3 in the denominator is the Laplace's correction for a 3-way classifier. 
The sentiment score is thus the mean of a discrete probability distribution and, as  \cite{GAMG16} put it, has \emph{``values of -1, 0 and +1 for negative, neutral and positive sentiment, respectively. 
The probabilities of each label are estimated from their relative frequencies, but when dealing with small samples (e.g., only a few tweets about a stock per day) it is recommended to estimate probabilities with Laplace estimate.''} 

\begin{table}[H]
\begin{tabular}{lllll}
\toprule
Created & Negative & Positive & Neutral & Compound\\
\midrule
2019-08-02 & 0.086543 & 0.223523 & 0.689960 & 0.208966 \\
2019-08-03 & 0.082495 & 0.249052 & 0.668458 & 0.237823 \\
2019-08-04 & 0.087113 & 0.247645 & 0.665240 & 0.232461 \\
2019-08-05 & 0.102785 & 0.236306 & 0.660908 & 0.192416 \\
2019-08-06 & 0.084345 & 0.245821 & 0.669837 & 0.235114 \\
\bottomrule
\end{tabular}
\caption{Aggregated Sentiment Scores Computed by VADER}
\label{tab:freq}
\end{table}

\section{Correlation Analysis and Causality}\label{sec:correlation}

\subsection{Correlation}
One of our project aims was to access how strongly (if at all) the changes in FTSE100 index returns and volatility, are correlated with the sentiment captured from the social media and financial news. 
We used Pearson’s correlation%
\footnote{\url{https://en.wikipedia.org/wiki/Pearson_correlation_coefficient}}, here denoted $r$, to access the level of correlation between sentiment of different data sets and stock market volatility and returns. 
The $r$ value varies between +1 (total linear correlation) and -1 (total linear anticorrelation) to indicate the strength and direction of the linear relationship between two variables.

\begin{table}[b]
\begin{tabular}{l|c|c|c|c}
\hline
\multicolumn{1}{c|}{Sentiment} & \multicolumn{2}{c|}{Returns} & \multicolumn{2}{c}{Returns-Next Day} \\ 
\cline{2-5} 
\multicolumn{1}{c|}{} & r & p-val. & r & \multicolumn{1}{c}{p-val.} \\ 
\hline
Negative & -0.1594 & 0.3757 & -0.1144 & 0.5330 \\
Positive & 0.2374 & 0.1834 &  0.1650 & 0.3669 \\
Neutral & -0.1277 & 0.4788 & -0.1500 & 0.4126 \\ 
\hline
\end{tabular}

\caption{Correlation results from tweets dataset - returns}
\label{tab:tweetercorr-ret}  
\end{table}

\begin{table}[b]
\begin{tabular}{l|c|c|c|c}
\hline
\multicolumn{1}{c|}{Sentiment} & \multicolumn{2}{c|}{Volatility} & \multicolumn{2}{c}{Volatility-Next Day} \\ 
\cline{2-5} 
\multicolumn{1}{c|}{} & r & p-val. & r & \multicolumn{1}{c}{p-val.} \\ 
\hline
Negative & -0.2051 & 0.2522 & -0.1646 & 0.3680 \\
Positive & -0.6979 & 0.0000 & -0.7009 & 0.0000 \\
Neutral & 0.7537 & 0.0000 & 0.7455 & 0.0000 \\ 
\hline
\end{tabular}

\caption{Correlation results from tweets dataset - volatility}
\label{tab:tweetercorr-vol}  
\end{table}

For p-values equal or below 0.05 (the significance level of 5\texttt{\%}) we can reject a null hypothesis and conclude that there is a statistically significant linear relationship between sentiment and stock market volatility and returns because the correlation coefficient is significantly different from zero.
Conversely, when the p-value is above 0.05 we can conclude there is not a significant linear relationship between sentiment scores and market volatility and returns as the correlation coefficient is not significantly different from zero.

\begin{figure}[htb]
  \centering
  \includegraphics[width=\linewidth]{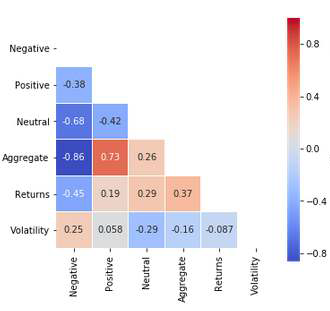}
  \caption{Correlation coefficients for headlines dataset - same day}
  \Description{Correlation coefficients}
    \label{fig:corrsameday}
\end{figure}

Negative, neutral, positive, and aggregate (compound) indicates sentiment, while returns means daily market returns and volatility shows stock market volatility. 
The heat map indicates correlation between sentiment on day t and next day’s market return and volatility measures. 
For instance, in Figure \ref{fig:corrsameday} the intersection between "Negative" on x-axis and "Returns" on y-axis indicates that r value is 0.45.

\begin{figure}[htb]
  \centering
  \includegraphics[width=\linewidth]{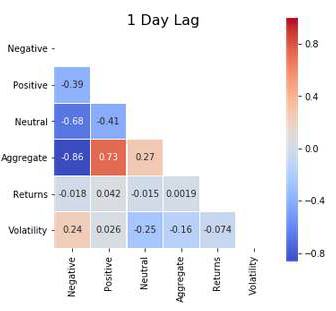}
  \caption{Correlation coefficients for headlines dataset - 1 day lag}
  \Description{Correlation coefficients}
  \label{fig:corr1day}
\end{figure}

The heatmap visualisation confirms the expectation of negative correlation between negative sentiment and stock market returns for a given day. 
Correlation coefficient denoted by r is equal to -0.45 and the p-value is below 0.05 (see \textbf{Table \ref{tab:corr-ret}:} headlines dataset), which means we can reject the null hypothesis and conclude that the relationship between negative sentiment captured from the headlines is moderate and statistically significant. 
We can also interpret this correlation as follows: if the sentiment of the headlines becomes increasingly negative, the closing price of the FTSE100 index at the end of day $t$ tends to be lower; when negative sentiment increases, returns decline and vice versa.

The aggregate sentiment score was obtained by Equation \ref{eq:sentd}; the higher the \emph{Sentd} value the stronger the positive sentiment and vice versa. 
An r-value of 0.37 indicates a weak correlation between aggregate sentiment and stock market returns. 
It suggests that if the average sentiment score increases the stock market returns will increase too.
Finally, if the average score decreases (becomes negative), the stock market returns would decrease as well.

Next, we tested if a sentiment at day t has a stronger impact on the stock market performance in the following day (t+1). 
To evaluate time-lag correlations between sentiment and stock market returns, we computed cross-correlation using a time lag of 1 day; the results are in Figure \ref{fig:corr1day}. 
They indicate that for a time lag of 1 day there is no statistically significant correlation between sentiment scores and market returns.
Also, there is a weak positive correlation between negative sentiment on one day and volatility the next day. 
An r-value of 0.24, with a p-value below 0.05, suggests that the variables (negative sentiment and volatility) are moving in tandem. 
I.e., if the negative sentiment on a given day increases then market volatility would also increase the next day.

\begin{figure}[htb]
    \centering
    \includegraphics[width=\linewidth]{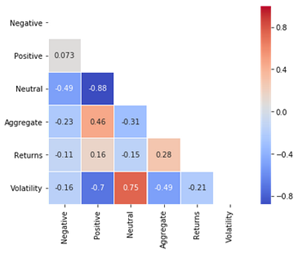}
    \caption{Correlation coefficients for tweets dataset - 1 day lag}
    \Description{Correlation coefficients}
    \label{fig:corrtweets-lag}
\end{figure}

As a further experiment, we tested the association between sentiment captured from tweets and stock market returns and volatility. 
Our findings are somewhat similar to those described in in the literature. 
As you can see from Figure \ref{fig:corrtweets-lag}, the low correlation coefficients indicate a weak correlation between positive, negative, neutral and aggregate sentiment scores and FTSE100 returns for a given day t. 
We obtained p-values above the 0.05 threshold; thus we conclude that no statistically-significant relationship is yet found between sentiment captured from tweets and FTSE100 returns.
Indeed the low correlation coefficients that we found are in line with the results obtained, e.g., by Mao et al. \cite{MCB11} and Nisar et al. \cite{NisYeu18}. 

Surprisingly, only the results obtained from tweet analysis indicate a strong correlation between sentiment (positive, neutral and average) and stock market volatility. 
For instance, a correlation coefficient of -0.7, and p-value below 0.05 (see \textbf{Table \ref{tab:tweetercorr-vol}}) indicates that there is a strong negative correlation between positive sentiment and the volatility of the market in the next day (t+1). 
It can be interpreted as follows: as the positive sentiment increases, market volatility decreases (two variables move in the opposite direction).

To summarize, the results above from different datasets suggest that the relationship between market sentiments and stock prices can be quite complex and may exist only in some of the time periods. 
It is unsurprising that the financial market exhibited different behaviours in different time periods.

Overall, our results from tweets dataset confirm the expectation of a negative correlation between positive sentiment and the volatility of the stock market, as the sentiment increases towards more positive the market volatility tend to decline as well: more positive news means calmer markets and less volatility. 

As the correlation coefficients (-0.45, 0.29 and 0.37) for headlines dataset are significantly less than 0.05 it can be concluded that the relationship between the sentiment (negative, neutral, aggregate, respectively) and the market returns at a given day t is relatively weak but still statistically significant. 
This means that the sentiment captured from headlines can be potentially used as a signal to predict the closing price of the FTSE100 index. 

For the 1-day time lag the correlation coefficients are very low and statistically insignificant in all three datasets.In conclusion, sentiment measured on a given day cannot be used to predict market returns on the next day. 
However, the opposite can be said about correlation between sentiment and market volatility. As the correlation coefficients (-0.70, 0.75 and -0.49) for tweets dataset are significantly lower than 0.05 it can be concluded that the relationship between the sentiment (positive, neutral, aggregate, respectively) and stock market volatility is relatively strong and statistically significant. Sentiment measured on a given day can be used to predict market volatility on the next day.

\subsection{Granger's Causality Test}
To verify whether market sentiment can indeed be useful for predicting FTSE100 Index movements, we decided to perform a Granger's causality test. 
This method is used to identify causality between two variables and whether one time series variable could be important in forecasting the other. 
In our case, we test whether the sentiment obtained from financial news and social media could be useful in forecasting the stock market performance and volatility.

\begin{equation}
  y_{t} = \alpha+\sum_{i=1}^k{\beta_{j}}{Y_{t-1}}+\sum_{j}^k{\lambda_{j}}{X_{t-j}} + \epsilon_{t}
  \label{eq:granger}
\end{equation}

If the p-value is less than 0.05, we could reject the null hypothesis and conclude that variable $X$ (sentiment) influence stock market changes and volatility.%

Granger's test provides more insights into how much predictive information one signal has about another one over a given lagged period. 
Here the p-value measures the  statistical significance of the causality between two variables (sentiment and market returns). 

Our causality testing found no reliable causality between the sentiment scores and the FTSE100 return in any lags. 
We found that causality slightly increased at a time lag of 2 days but it remained statistically insignificant.

It is also possible to test whether returns can be said to affect sentiment.
For such case Granger's test indicated that is it more likely that the market returns cause negative sentiment: the p-value is below the significance threshold of 0.05.

To summarize, the Granger's causality analysis of three different datasets (headlines, news stories and tweets) with FTSE returns and volatility has shown that, in general, sentiment obtained from news or social media was found to ``cause'' neither changes to the FTSE100 index closing prices nor changes in market volatility. 
The p-values were all above the significance threshold, which means our null hypothesis could not be rejected.

\section{Latent Dirichlet allocation}\label{sec:lda}

The most commonly-used method for topic modelling, or topic discovery from a large number of documents, is Latent Dirichlet allocation (LDA). 
LDA is a generative topic model which generates combination of latent topics from a collection of documents, where each combination of topics produces words from the collection’s vocabulary with certain probabilities. 
The process of running LDA consists of several steps. 
A distribution on topics is first sampled from a Dirichlet distribution, and a topic is further chosen based on this distribution. 
Moreover, each document is modelled as a distribution over topics, and a topic is represented as a distribution over words. 

LDA allows a set of news stories and tweets to be categorized into their underlying topics. 
According to Atkins et al. \cite{ANG18} “a topic is a set of words, where each word has a probability of appearance in documents labelled with the topic. 
Each document is a mixture of corpus-wide topics, and each word is drawn from one of these topics. 
From a high-level, topic modelling extrapolates backwards from a set of documents to infer the topics that could have generated them – hence the generative model”. 
Although LDA reduces the dimensionality of the data by producing a small number of topics, it is relatively computationally heavy (yet polynomial time $O(n^k)$). 

Recently \cite{ANG18} proposed an LDA model to represent information from news sources and then used a simple Na\:ive Bayes classifier to predict the direction of the market volatility. 
The results indicate 56\texttt{\%} accuracy in predicting directional stock market volatility on the arrival of new information. 
The authors concluded that “volatility movements are more predictable than asset price movements when using financial news as machine learning input, and hence could potentially be exploited in pricing derivatives contracts via quantifying volatility”.

Others, Mahajan et al. \cite{MDH08} also used LDA to identify topics of financial news and then to predict a rise or fall in the stock markets based on topics extracted from financial news. 
Their developed classifier achieved 60\texttt{\%} accuracy in predicting market direction.

We have followed Atkins' methodology and to assess whether topics extracted from tweets and news headlines can be used to predict directional changes in market volatility. 
Let us now see the steps we followed to perform LDA and use its produced topic distribution to predict next day’s market volatility (‘UP’ or ‘DOWN’).

\subsection{Feature Vectors}
We followed Kelechava’s methodology which can be found on its GitHub site%
\footnote{Please see \url{https://github.com/marcmuon/nlp_yelp_review_unsupervised}} %
to convert topics into feature vectors. Then, an LDA model was used to get the distribution of 15 topics for every day's headlines. 
This 15-dimensional vector will be used later as a feature vector for a classification problem, to assess if topics obtained on a certain day can be used to predict the direction of market volatility the next day.

The feature vector for an interval is a topic-count sparse vector, representing the number of times each topic appears in headlines/tweets or articles within the interval. 
Some topics may appear more than once, and some not at all. 
The target vector is then constructed by pairing binary direction labels from market volatility data to each feature vector. 
For instance, we are using headlines from day t to predict the direction of movement (increase/decrease) of volatility over the next day $t+1$.

\subsection{Applying the classifier to unseen test sets}
Thanks to the preparation described earlier, we could build an LDA model and train our classifier. 
However, in order to access our model performance, we tested our model by computing a feature vectors from unseen test data and running a simple logistic regression model to predict if the next day’s market volatility will increase or decrease. 

\begin{figure}[ht]
  \centering
  \includegraphics[width=\linewidth]{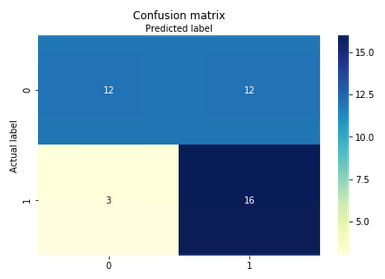}
  \caption{Confusion matrix - headlines dataset}
\end{figure}

Our model applied to the headlines dataset obtained an accuracy of 65\%. 
It indicates that topics extracted from news could be used as a signal to predict direction of market volatility next day. 
The results obtained from our modes are very similar to the ones of Atkins et al. \cite{ANG18} and Mahajan et al. \cite{MDH08}.
The accuracy was slightly lower for tweets dataset, which can be explained by the fact that tweets text typically contains abbreviations, emojis and grammatical errors which could make it harder to capture topics from tweets. 

\subsection{Visualization of the emergent topics}

\begin{figure}[htb]
  \centering
  \includegraphics[width=\linewidth]{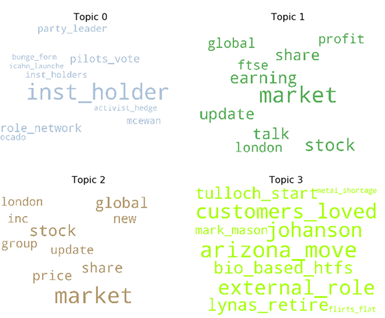}
  \caption{LDA topic modeling of our headlines corpus}
    \label{fig:topicscloud}
\end{figure}

The ``cloud'' presented in Figure \ref{fig:topicscloud} was obtained from headlines dataset. 
Each topic contains a maximum of 10 words. 
It is interesting to notice that topics captured from headlines news are very different from topics obtained from the news stories. 

As each dataset contains slightly different topics and key words, it would be interesting to assess whether a combination of three different datasets could help to improve the prediction of our model (in the hope that the datasets would ``complement'' each other).

\section{Results}\label{sec:results}

Financial markets are influenced by a number of quantitative factors, ranging from company announcements and performance indicators such as EBITDA, to sentiment captured from social media and financial news. 
As described in Section \ref{sec:related} several studies have modeled and tested the association between ``signals''---sentiment--- from the news and market performance. 
To evaluate our own sentiment extraction we have applied Pearson’s correlation coefficient to quantify the level of correlation between sentiment of our data collection and stock market volatility and returns. 

Tables \ref{tab:corr-ret}---\ref{tab:corr-vol} below summarise the results of the correlation analysis. 
The findings suggest that there is evidence of a weak correlation between sentiment captured from headlines and FTSE100 returns the same day. 
However, the correlation between sentiment on a given day and market returns the next day is not statistically significant.

\begin{table*}[htb]
\centering
\begin{tabular}{c|c|c|c|c|c|c|c|c|c|c|c|c}
\hline
Sentiment & \multicolumn{4}{c|}{Headlines} & \multicolumn{4}{c|}{Tweets} & \multicolumn{4}{c}{News Stories} \\ \cline{2-13}
 & \multicolumn{2}{c|}{Same day} & \multicolumn{2}{c|}{1 day lag} & \multicolumn{2}{c|}{Same day} & \multicolumn{2}{c|}{1 day lag} & \multicolumn{2}{c|}{Same day} & \multicolumn{2}{c}{1 day lag} \\ 
 \cline{2-13}
 & r & p-value & r & p-value & r & p-value & r & p-value & r & p-value & r & p-value \\ 
 \hline
Negative & -0.4506 & 0.0000 & -0.0180 & 0.8327 & -0.1594 & 0.3757 & -0.1144 & 0.5330 & 0.2539 & 0.1924 & -0.0823 & 0.6832 \\
Positive & 0.1900 & 0.0235 & 0.0422 & 0.6194 & 0.2374 & 0.1834 & 0.1650 & 0.3669 & -0.0209 & 0.9160 & 0.0354 & 0.8607 \\
Neutral & 0.2912 & 0.0044 & -0.0151 & 0.8590 & -0.1277 & 0.4788 & -0.1500 & 0.4126 & -0.1140 & 0.5635 & 0.0077 & 0.9679 \\
Aggregate & 0.3671 & 0.0000 & 0.0019 & 0.9820 & 0.1540 & 0.3922 & 0.2842 & 0.1148 & -0.1505 & 0.4400 & 0.0267 & 0.8947 \\ 
\hline
\end{tabular}
\caption{Correlation between sentiment and stock market returns}
\label{tab:corr-ret}
\end{table*}

\begin{table*}[htb]
\centering
\begin{tabular}{c|c|c|c|c|c|c|c|c|c|c|c|c}
\hline
Sentiment & \multicolumn{4}{c|}{Headlines} & \multicolumn{4}{c|}{Tweets} & \multicolumn{4}{c}{News Stories} \\ \cline{2-13}
 & \multicolumn{2}{c|}{Same day} & \multicolumn{2}{c|}{1 day lag} & \multicolumn{2}{c|}{Same day} & \multicolumn{2}{c|}{1 day lag} & \multicolumn{2}{c|}{Same day} & \multicolumn{2}{c}{1 day lag} \\ \cline{2-13}
 & r & p-value & r & p-value & r & p-value & r & p-value & r & p-value & r & p-value \\ \hline
Negative & 0.2492 & 0.0028 & 0.2350 & 0.0050 & -0.2051 & 0.2522 & -0.1646 & 0.3680 & 0.2425 & 0.2137 & 0.3615 & 0.0639 \\
Positive & 0.0583 & 0.4904 & 0.0262 & 0.7573 & -0.6979 & 0.0000 & -0.7009 & 0.0000 & 0.0060 & 0.9760 & -0.0970 & 0.6304 \\
Neutral & -0.2918 & 0.0042 & -0.2541 & 0.0024 & 0.7537 & 0.0000 & 0.7455 & 0.0000 & -0.1374 & 0.4857 & -0.0999 & 0.6201 \\
Aggregate & -0.1611 & 0.0555 & -0.1609 & 0.0566 & -0.4873 & 0.0040 & -0.4922 & 0.0042 & -0.0640 & 0.7463 & -0.2207 & 0.2685 \\ \hline
\end{tabular}
\caption{Correlation between sentiment and stock market volatility}
\label{tab:corr-vol}
\end{table*}

As shown in Table \ref{tab:corr-vol}, of the correlation analysis, an increase in positive sentiment captured from tweets leads to decreased market volatility (strong negative correlation between positive sentiment and market volatility). 
For instance, a -0.7 correlation between day’s t positive sentiment and the next day’s market volatility is statistically significant, meaning that if the tweets show an increasingly positive sentiment, FTSE100 volatility decreases.

The correlation between average sentiment scores and next day’s volatility measures is strongly negative and statically significant too. 
Hence we reject the null hypothesis and conclude that positive and average sentiment scores calculated from tweets can be used to predict next day’s market volatility. 
An important additional suggestion is that that an increase of the average, i.e., when it tends to 1 (increasing positive sentiment), can be associated to  calmer, less volatile markets.  

The strongest correlation between sentiment and volatility measures was detected in our tweets dataset, while no correlation or weak correlation was found in headlines and news stories dataset. 
This can be potentially explained by the timelines of tweets. 
Twitter users, analysts as well as financial companies can express their options via Tweeter much faster and-not to be overlooked-they can time the publication of their tweets.
The process to publish news and analyses is typically longer. 
Contents need to be gathered, selected and commented by journalists and then proofread by an editorial team; this often leads to 1 to 2 days before a news story is released on a professional newslet.

It is important to mention, that some of our findings are aligned to the results already available in the literature as low (and statistically-insignificant) correlations between sentiment captured from tweets and stock markets were obtained by several previous studies.
Our project went further to assess whether topics derived from financial news and social media have greater accuracy in predicting market volatility. 
To do so, we built an LDA model to extract feature vectors from each day’s news and then used a logistic regression to predict the direction of market volatility the next day.
To measure our classifier performance, we used measures such as accuracy, recall, precision, and F1 score. 
All these measures were obtained using the well-known Python Scikit-lean libraries/packages%
\footnote{Please see \url{https://scikit-learn.org/}}.

Table \ref{tab:summary} summarises the detailed results of our LDA and classification model. 
All three models produced somewhat similar results which are in line with previous studies such as Atkins et al. \cite{ANG18}  and Mahajan et al. \cite{MDH08}.
Despite the fact that the language used in tweets is informal, filled with acronyms and sometimes errors, the results we obtained from our Tweeter datasets were surprisingly good, with an accuracy that almost matches that obtained from the headlines dataset.

\begin{table}[htb]
  \label{tab:commands}
  \begin{tabular}{cclll}
    \toprule
    Dataset & Accuracy & Recall & Precision & F1 score\\
    \midrule
    \texttt{{Headlines}} & 0.65& 0.65 & 0.64 & 0.64 \\
    \texttt{{Tweets}}& 0.64& 0.65 & 0.70 & 0.64 \\
    \texttt{{Stories}}& 0.67& 0.67 & 0.81 & 0.63 \\
    \bottomrule
  \end{tabular}
  \caption{Summary of the Results}
  \label{tab:summary}
\end{table}

\section{Conclusion and Future Work}\label{sec:conclusion}
Our project involved performing a correlation analysis to compare daily sentiment with daily changes in FTSE100 returns and volatility. 
Overall, correlation analysis shows that sentiment captured from headlines could be used as a signal to predict market returns, but not so much volatility. 
The opposite was true for our tweets dataset. 
A correlation coefficient of -0.7, and p-value below 0.05 indicated that there is a strong negative correlation between positive sentiment captured from the tweets and the volatility of the market next day (t+1). 
It suggests that as the positive sentiment increases, market volatility decreases (the two variables move in the opposite direction).

Of the three different data sets that were created, the most promising results were obtained from the headlines data set; this can be explained by the fact that this data set was the largest and had the longest time series. 
It would be beneficial to expand the correlation analysis by building a larger data corpus.

In addition, we observed a slightly stronger correlation between sentiment captured from tweets containing cashtags \texttt{(\$)} and market returns compared with tweets containing only hashtags \texttt{(\#)} or multiple keywords. 
When it comes to building a tweets data set, there are some issues associated with hashtags or keywords. 
Many tweets will contain multiple keywords, but only actually express an emotion towards one of them. 
Using more advanced natural language processing techniques to identify the subject of a tweet could potentially help reduce noise in Twitter data.  

Results obtained with Granger's test indicate that, in general, sentiment obtained from news and social media does not seem to ``cause'' either changes in FTSE100 index prices or the volatility of the index; all p-values obtained in the tests where above 0.10 threshold so the null hypothesis could not be rejected.

The topics extracted from news sources can be used in predicting directional market volatility. 
It is surprising that topics alone contain a valuable information that can be used to predict the direction of market volatility. 

The evaluation of the classification model has demonstrated good prediction accuracy. 
Our model applied to the headlines dataset obtained an accuracy of 65\%. 
It indicates that topics extracted from news could be used as a signal to predict the direction of market volatility next day. 
It was noticed that the accuracy of the model tends to depend on the number of topics chosen. 
There are different techniques that could be used to select an optimal number of topics; however, some of them, especially the development of high-frequency LDA models, are computationally expensive and would require a preliminary scalability analysis and a capable architecture to run on.

The future work could include building a proprietary sentiment scoring system or a system that detects mood from the news (Calm, Alert, Sure, Vital, Kind, and Happy). 
Previous work by Bollen et al. \cite{Bol11} indicated that mood captured from tweets can help to predict the direction of Down Jones index with 86.7\% accuracy. 
However, it would be interesting to see if this model could be improved by using a larger corpus of headlines and news stories, instead of tweets only.

Finally, we acknowledge that one of the key limitations of this research was a relatively small sample size. 
In order to obtain more reliable results we believe a larger dataset is necessary.

\bibliographystyle{ACM-Reference-Format}
\bibliography{senti+topic}

\end{document}